\documentclass[fleqn,twoside]{article}%
\usepackage{amsmath}
\usepackage{amsfonts}
\usepackage{amssymb}
\usepackage{graphicx}%
\def\be{\begin{equation}}
\def\ee{\end{equation}}
\def\bi{\bibitem}

\begin{document}

\title{Transient Crossing of Phantom divide line $w_{\Lambda}=-1$ under Gauss-Bonnet interaction}
\author{Abhik Kumar Sanyal}
\maketitle
\begin{center}
Dept. of Physics, Jangipur College, Murshidabad, \noindent
West Bengal, India - 742213. \\

\end{center}
\noindent
\begin{abstract}
Smooth double crossing of the phantom barrier $w_{\Lambda} = -1$ has been found possible in cosmological model
with Gauss-Bonnet-scalar interaction, in the presence of background cold dark matter. Such crossing has been
observed to be a sufficiently late time phenomena and independent of the sign of Gauss-Bonnet-scalar
interaction. The luminosity distance versus redshift curve shows a perfect fit with the $\Lambda CDM$ model up
to $z=3.5$.
\end{abstract}
\noindent
\section{Introduction}
The puzzle associated with recent cosmic acceleration, triggered by $70\%$ of dark energy or more \cite{a1} is
far from being resolved uniquely. In the mean time, cosmologists are being confronted with yet another more
intriguing challenge to explain the crossing of the so called phantom divide line $(w_{\Lambda}=-1)$, at
sufficiently late time of cosmological evolution. Some recent analysis \cite{a1},\cite{b1} of the presently
available observational data are in favour of the value $w_{de}<-1$, at present., $w_{de}$ being the dark energy
equation of state. There are also a lot of evidence all around \cite{c}, of a dynamical dark energy equation of
state, which has crossed the so called phantom divide line $w_{\Lambda}=-1$ recently, at the value of red-shift
parameter $z \approx 0.2$. Apparently though the problem turns out to be more serious and complicated, but then,
the puzzle of crossing the phantom divide line has also rendered some sort of selection rule. $\Lambda
CDM$-model, which is known to suffer from the disease of fine tuning (see \cite{d} for a comprehensive review)
can now be ruled out due to the requirement of a dynamic state parameter. Further, if the analysis of Vikman
\cite{e} is correct, then it is not possible to cross the phantom divide line in a single minimally coupled
scalar field theory, without violating the stability both at the classical \cite{f} and also at the quantum
mechanical levels \cite{g}, (though it has recently been inferred \cite{on} that quantum Effects which induce
the $w<-1$ phase, are stable in the $\phi^4$ model). Thus single minimally coupled scalar field models like
quintessence $(w>-1)$ and phantom $(w<-1)$ are to be kept aside. Consequently, we are now left with some what
more complicated models. One of these is a hybrid model, composed of two scalar fields, viz, quintessence and
phantom - usually dubbed as quintom model \cite{h}. Other models like non-minimal scalar tensor theory of
gravity \cite{i}, hessence \cite{j} and models including higher order curvature invariant terms
\cite{k} also exist in the literature. \\
Gauss-Bonnet term is yet another candidate which may be pursued for the purpose. The possibility of crossing the
phantom divide line through Gauss-Bonnet interaction has been explored in some recent works \cite{l},\cite{m}.
But then, these models are even complicated in the sense that either brane-world scenario \cite{l} or scalar
field and matter coupling \cite{m} are invoked. In this article the possibility of smooth crossing of the
phantom divide line $w_{\Lambda}=-1$ has been expatiated simply by introducing Gauss-Bonnet-Scalar coupling term
in the
4-dimensional Einstein-Hilbert action.\\
Gauss-Bonnet term arises naturally as the leading order of the $\alpha'$ expansion of heterotic superstring
theory, where, $\alpha'$ is the inverse string tension \cite{n}. Gauss-Bonnet term is topologically invariant
and thus does not contribute to the field equations in four dimensions. However, the low energy limit of the
string theory gives rise to the dilatonic scalar field which is found to be coupled with various curvature
invariant terms \cite{o}. The leading quadratic correction gives rise to Gauss-Bonnet term with a dilatonic
coupling \cite{p}. Therefore it is reasonable to consider Gauss-Bonnet interaction in four dimension with
dilatonic-scalar coupling. Several works with Gauss-Bonnet-dilatonic coupling are already present in the
literature \cite{q}. In particular, important issues like - late time dominance of dark energy after a scaling
matter era and thus alleviating the coincidence problem, crossing the phantom divide line and compatibility with
the observed spectrum of cosmic background radiation have also been addressed recently \cite{km}.\\
In a recent work with Gauss-Bonnet interaction \cite{a}, a solution in the form $a=a_{0}e^{A\sqrt t}$ ($a$ being
the scale factor, and $A>0$) has is been found to satisfy the field equations with different forms (sum of
exponentials, sum of inverse exponentials, sum of powers and even quadratic) of potentials. Solution in a more
general form $(a=a_{0}e^{A t^f}), A>0, 0<f<1$, for Einstein's gravity with a minimally coupled scalar field was
found in the nineties \cite{r} and was dubbed as intermediate inflation. We \cite{a}, on the other hand,
observed that such solution depicts a transition from decelerated to accelerated expansion at sufficiently later
epoch of cosmic evolution, which asymptotically goes over to de-Sitter expansion. Thus, it appeared that such
solution may construct viable cosmological models of present interests. Under this consequence, a comprehensive
analysis has been carried out \cite{b} with such solution in the context of a generalized k-essence model. It
has been observed that it admits scaling solution with a natural exit from it at a later epoch of cosmic
evolution, leading to late time acceleration with asymptotic de-Sitter expansion. The corresponding scalar field
has also been found to behave as a tracker field \cite{s}, thus avoiding cosmic coincidence problem. \\
In the present work, we show that Gauss-Bonnet-Dilatonic scalar coupling with Einstein's gravity in four
dimensions, admits solution in a general form $(a=a_{0}e^{A t^f}), A>0, 0<f<1$, which is viable of crossing the
phantom divide line twice, once from above and the other from below in the recent epoch. Since the crossing is
transient, so we may conclude that it does not show any pathological behaviour like Big-Rip \cite{f}, at least
in the classical level.

\section{The Model with Gauss-Bonnet Interaction}

We start with the following action containing Gauss-Bonnet
interaction

\be S=\int d^4x\sqrt{-g}[\frac{R}{2\kappa^2}+
\frac{\Lambda(\phi)}{8}G(R)-\frac{1}{2}\phi,_{\mu}\phi'^{\mu}-V(\phi)+L_{m}],
\ee
where,
\[G(R)=R^2-4R_{\mu\nu}R^{\mu\nu}+R_{\mu\nu\rho\sigma}R^{\mu\nu\rho\sigma}\]
is the Gauss-Bonnet term which appears in the action with a
coupling parameter $\Lambda(\phi)$ and $L_{m}$ is the matter
Lagrangian. For the spatially flat Robertson-Walker space-time
\[ds^2=-dt^2+a^2(t)[dr^2+r^2 d\theta^2+r^2 sin^2\theta d\phi^2],\]
the field equations in terms of the Hubble parameter $H=\frac{\dot
a}{a}$, are

\be 2\dot H+3H^2=-[\frac{1}{2}\dot\phi^2-V(\phi)+2\Lambda'\dot\phi(H\dot
H+H^3)+(\Lambda'\ddot\phi+\Lambda''\dot\phi^2)H^2+p_{m}]=- (p_{de}+p_{m}), \ee

\be 3H^2=[\frac{1}{2}\dot \phi^2+V(\phi)-3\Lambda'\dot{\phi}H^3+\rho_{m} ]= (\rho_{de}+ \rho_{m}), \ee in the
units $\kappa^2 (= 8\pi G) = \hbar = c =1$. In our analysis the Gauss-Bonnet scalar interaction plays the role
of dark energy, for which suffix ($de$) has been introduced. Thus, $p_{de}$ and $\rho_{de}$ are the effective
pressure and the energy density generated by the Gauss-Bonnet-scalar interaction, while $p_{m}$ and $\rho_{m}$
are the pressure and the energy density corresponding to background matter distribution respectively. The
background matter satisfies the equation of state,

\be \rho_{m}=\rho_{i}a^{-3(1+w_{m})}, \ee where, $\rho_{i}$ is a constant and $w_{m}$ is the state parameter of
the background matter. In addition we have got the $\phi$ variation equation

\[ (\ddot\phi+3H\dot\phi+V')=3\Lambda'H^2(\dot H+H^2), \] which is
not an independent equation and will not be required in our analysis. In the above, over-dot and dash ($\prime$)
stand for differentiations with respect to the proper time $t$ and $\phi$ respectively. Now, in view of
equations (2) through (4), we are required to solve for $a, \phi, V(\phi), \Lambda(\phi), p_{m}$ and $\rho_{m}$,
which requires three additional assumptions. Firstly, we consider that the Universe is filled with cold dark
background matter with equation of state, $p_{m} = 0$ while the second assumption is the one made previously in
\cite{a}, viz.,

\be \Lambda'\dot\phi= \lambda, \ee where, $\lambda$ is a constant.
This, as indicated in \cite{a} is physically reasonable, since it
implies that the Gauss-Bonnet coupling parameter $\Lambda(\phi(t))
= \lambda t$, grows in time to contribute at the later epoch of
cosmological evolution. In view of the above assumption the field
equations (2) through (4) are expressed as,

\be 2\dot H+3H^2=-[\frac{1}{2}\dot \phi^2-V(\phi)+2\lambda H\dot H+2\lambda H^3]=- p_{de}, \ee

\be 3H^2=[\frac{1}{2}\dot \phi^2+V(\phi)-\lambda H^3+\rho_{m}]=(\rho_{de}+\rho_{m}), \ee and,

\be\rho_{m}=\rho_{i} a^{-3}.\ee Now, for our third assumption, we start from the ansatz,

\be H=\frac{f}{nt^{1-f}},\ee with $0 < f < 1, n = A^{-1} >0$, which leads to the form of the solution of the
scale factor mentioned in the introduction. Thus the complete set of solutions are given by,

\[a=a_{0}\exp(\frac{t^f}{n})
;~~\rho_{m}=\frac{\rho_{i}}{a_{0}^3}\exp(-\frac{3}{n}t^f);~~p_{de}=\frac{2f(1-f)}{nt^{(2-f)}}
-\frac{3f^2}{n^2t^{2(1-f)}};~~\rho_{de}=\frac{3f^2}{n^2 t^{2(1-f)}}-\frac{\rho_{i}}{a_{0}^3
\exp{(\frac{3}{n}t^f})};\]
\[w_{de}= a_{0}^3\left(\frac{2nf(1-f)t^{-f}-3f^2}{3a_{0}^3f^2-
\rho_{i}n^2t^{2(1-f)}\exp(-\frac{3}{n}t^f)}\right);\] \be\rho_{de}+3p_{de}= \frac{6f(1-f)}{ n
t^{(2-f)}}-\frac{6f^2}{n^2t^{2(1-f)}} -\frac{\rho_{i}}{a_{0}^3\exp(\frac{3}{n}t^f)};~~
\rho_{de}+p_{de}=\frac{2f(1-f)}{ n t^{2-f}}-\frac{\rho_{i}}{a_{0}^3\exp(\frac{3}{n}t^f)};~~ \ee
\[\dot\phi^2=\frac{\lambda f^3}{n^3 t^{3(1-f)}}+
\frac{2\lambda f^2(1-f)}{n^2 t^{3-2f}}+\frac{2f(1-f)}{n t^{2-f}}- \frac{\rho_{i}}{a_{o}^3
\exp{(\frac{3}{n}t^f})},\]
\[V=\frac{3\lambda f^3}{2n^3 t^{3(1-f)}}-\frac{\lambda
f^2(1-f)}{n^2 t^{3-2f}}+\frac{3f^2}{n^2 t^{2(1-f)}}-\frac{f(1-f)}{n t^{2-f}}- \frac{\rho_{i}}{2a_{o}^3
\exp{(\frac{3}{n}t^f})}.\] Above set of solutions (10) indicates that such a model of the Universe admits an
early deceleration, but during evolution it starts accelerating since strong energy condition is violated,
$\rho_{de}+3p_{de}<0$. Further, the dark energy equation of state also admits the possibility of crossing the
$w_{\Lambda} = -1$ line, since, transient violation of the weak energy condition, $\rho_{de}+p_{de} <  0$ is
seemingly possible. Finally, the equation of state asymptotically touches the $w_{\Lambda} = -1$ line from above
and behaves as cosmological constant. To show such behavior graphically, let us express the state parameter
$w_{de}$ in terms of the red-shift parameter $z$ which is defined as,

\[1+ z=\frac{a(t_{o})}{a(t)}=\exp[{\frac{1}{n}(t_{o}^f-t^f)}],\]
where, $a(t_{o})$ is the present value of the scale factor, while
$a(t)$ is that value at some arbitrary time $t$, when the light
was emitted from a cosmological source. Thus,

\be t^f=t_{o}^f - n \ln(1+z).\ee In view of equation (11) $w_{de}$ can be expressed as,

\be w_{de}= \left(\frac{(2nf(1-f)-3f^2[t_{o}^{f}-n\ln{(1+z)}]}{3f^2[t_{o}^{f}-n\ln{(1+z)}] - \rho_{i}
n^2[t_{o}^{f}-n\ln{(1+z)}]^{\frac{2-f}{f}} \exp{(-\frac{3}{n}[t_{o}^{f}-n\ln{(1+z)}])}}\right),\ee where,
$a_{o}$ has been set equal to one without any loss of generality. Let us now choose $f=0.5$. The motivation of
choosing such a value of $f$ is twofold. Primarily, it is impossible to find an explicit form of the potential
$V = V(\phi)$, otherwise. Further, since $n$ has the dimension of $t^f$, so the parameter $n^2$ gets a
comfortable dimension of time. If we now take up some more numbers, like the present value of the Hubble
parameter $H_{o}^{-1}$ and the age of the Universe $t_{o}$ as,

\[H_{0}^{-1}= \frac{9.78}{h}~Gyr, ~~ t_{0}=13~Gyr,\]
then, for $h=0.66$, $n$ can be found from the ansatz (9) as $n=2.0552$. Further taking the present value of the
matter density parameter $\Omega_{mo}= 0.26$, we find in view of solution (10),
\[\Omega_{mo}=  \frac{\rho_{mo}}{\rho_{co}}=
 \rho_{i}(\frac{H_{o}^{-2}\exp{(-\frac{3}{n}t_{o}^f})}{3})=0.26,\] where, $\rho_{mo}$ and $\rho_{co}$ are the
present values of the matter density and the critical density respectively. Thus, we find,

\[n^2 \rho_{i}= 2.897.\]
Noting that in this model the red-shift parameter does not go beyond the value $z=4.78$, we plot the dark energy
equation of state parameter $(w_{de})$ versus the red-shift parameter $(z)$ in figure (1). It is apparent that
the phantom divide line $\omega_{\Lambda}$ has been crossed twice, once from above at $z \approx 1.92$ and then
from below at $z \approx 0.39$. Such double crossing of the phantom divide line is devoid of any sort of
pathological behaviour.

\begin{figure}
[ptb]
\begin{center}
\includegraphics[
height=2.034in, width=3.3797in] {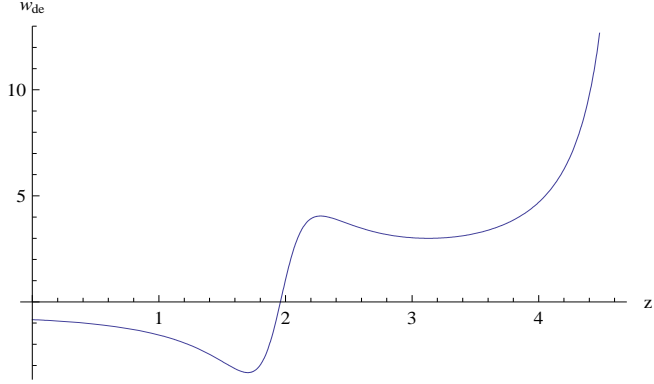} \caption{State parameters $w_{de}(z)$ has been plotted against the
red-shift parameter $z$, (with, $f=0.5, h=0.66, t_{0}=13~Gyr, \Omega_{mo}=0.26$). Smooth double crossing of the
Cosmological constant barrier is observed at sufficiently later epoch, $z\approx 1.92$ from above and $z\approx
0.39$ from below.}
\end{center}
\end{figure}
To check how far our present model fits with the standard $\Lambda CDM$ model, we also make the
luminosity-redshift and distance modulus-redshift plots. For $\Lambda CDM$ model the relation between the
luminosity distance and redshift is the following,
\[H_{0}dL=(1+z)\int_{0}^{z} \frac{dz}{0.74+0.26(1+z)^3},\]
while the relation corresponding to the present model is,
\[H_{0}dL=\frac{1+z}{\sqrt t_{0}}\int_{0}^{z}[\sqrt t_{0}-n \ln{(1+z)}]dz,\]
with $t_{0} = 13$ Gyr., and $n = 2.055$. The plot (figure 2) shows a perfect fit between the two models up to $z
= 3.5$. There is a little discrepancy  there after. Since, luminosity distance has already been expressed as a
function of redshift, so the relation between distance modulus and redshift may be found in view of the
following equation,
\[m-M=5\log _{10}(\frac{dL}{Mpc})+25,\]
where, $m$ and $M$ are the apparent and absolute bolometric magnitudes respectively. However, since we use
$H_{0}dl$ instead, so our relation is slightly modified as,
\[m-M=5\log _{10}(DL)+31,\]
where, $DL = H_{0} dL$. The plot (figure 3) demonstrates that the two models are practically indistinguishable.

\begin{figure}
[ptb]
\begin{center}
\includegraphics[
height=2.034in, width=3.3797in] {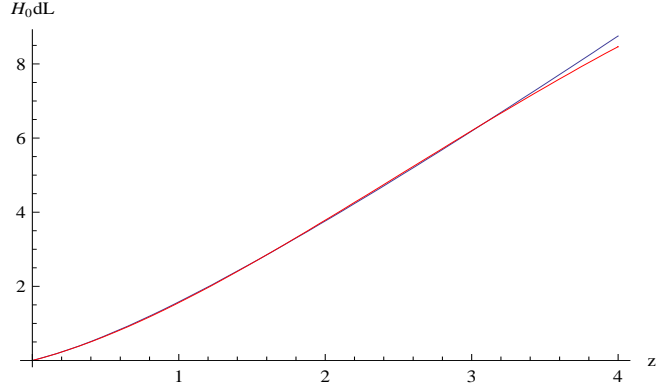} \caption{The fit is almost perfect up to $z = 3.5$. A little
discrepancy is observed there after as $\Lambda CDM$ model (blue) slightly takes over the present model (red).}
\end{center}
\end{figure}

\begin{figure}
[ptb]
\begin{center}
\includegraphics[
height=2.034in, width=3.3797in] {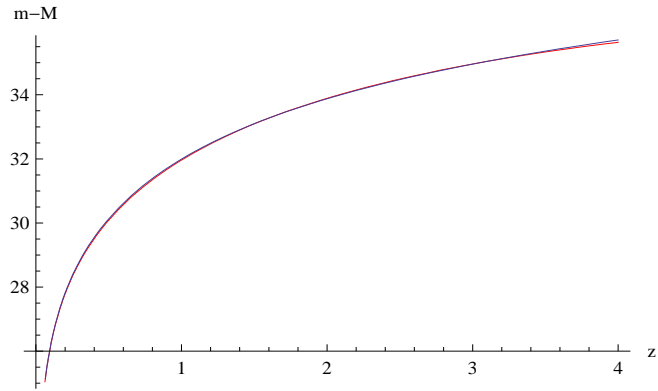} \caption{The fit is absolutely perfect and the two models (blue for
$\Lambda CDM$ and red for the present) are practically indistinguishable.}
\end{center}
\end{figure}
Now we can proceed to make some even more comfortable choice of the parameters of the theory, like $f=0.5$, and
$t_{0} = 13$ Gyr. as before, but with $n=2$, for which $H_{0}^{-1}=14.42$ Gyr., in view of ansatz (9), which
corresponds to $h\approx 0.68$. The scale factor now has a convenient form as $a = \exp{\frac{\sqrt{t}}{2}}$.
With these values one can find,
\[n^2 \rho_{i}= 3.09,\]
for $\Omega_{mo} = 0.24$. The plot (fig.4) is almost the same as before with transient double crossing. Other
plots viz., luminosity distance versus redshift (in figure 5) and distance modulus versus redshift (in figure 6)
show even better fit than the earlier one, with the standard $\Lambda CDM$ model.

\begin{figure}
[ptb]
\begin{center}
\includegraphics[
height=2.034in, width=3.3797in] {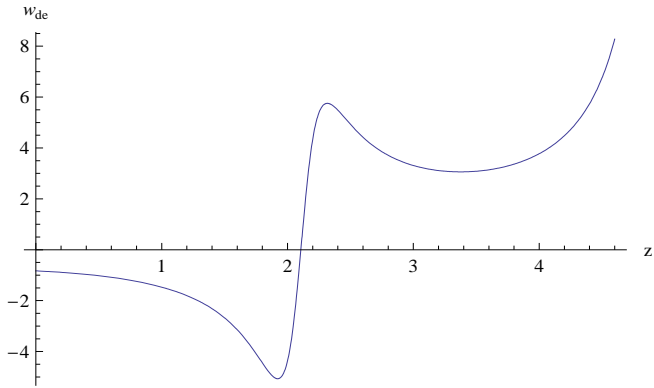} \caption{State parameters $w_{de}(z)$ has been plotted against the
red-shift parameter $z$, (with, $f = 0.5, h \approx 0.68, t_{0} = 13~Gyr., \Omega_{mo} = 0.24$). Smooth double
crossing of the Cosmological constant barrier is observed at sufficiently later epoch, $z\approx 2.085$ from
above and $z\approx 0.46$ from below.}
\end{center}
\end{figure}

\begin{figure}
[ptb]
\begin{center}
\includegraphics[
height=2.034in, width=3.3797in] {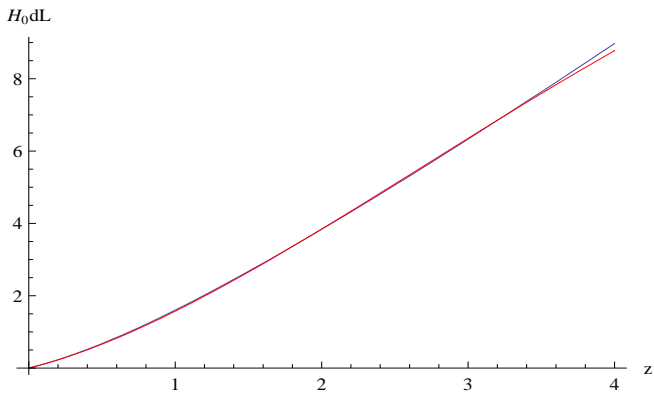} \caption{The fit is perfect up-to $z = 3.6$. The $\Lambda CDM$ model
(blue) slightly overtakes the present model (red) there after.}
\end{center}
\end{figure}

\begin{figure}
[ptb]
\begin{center}
\includegraphics[
height=2.034in, width=3.3797in] {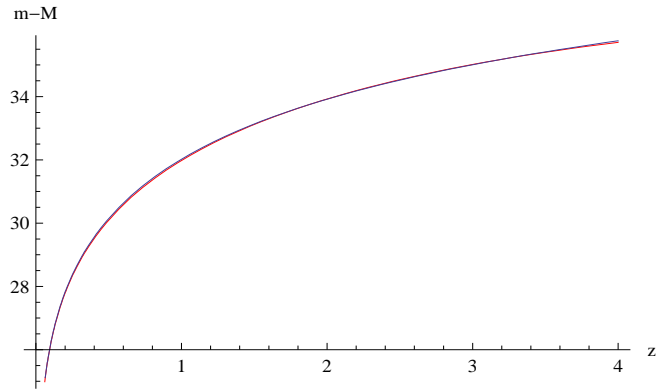} \caption{The fit is absolutely perfect and there is practically no way
to distinguish the $\Lambda CDM$ model (blue) with the present one (red).}
\end{center}
\end{figure}
Thus, we observe that with the age $t_{0} = 13$ Gyr., and $0.66 \le h \le 0.68$, such transient crossing of the
phantom divide line is permissible for the present value of dark energy density parameter
$\Omega_{de}|_{present} \ge 0.74$. In order to consider some higher value of the age of the Universe, $t_{0} =
13.73$ Gyr. (say) as suggested by Spergel et al\cite{t}, either one has to go to almost the lowest limiting
value of $h \approx 0.61$ \cite{u} or one has to accept much higher value of the present dark energy density
parameter $\Omega_{de}|_{present} > 0.78$, otherwise, the state parameter versus redshift plot shows certain
discontinuities. We certainly remember that in order to simplify the field equations considerably, we have made
one important assumption, viz., in equation (5). Relaxing this assumption one might get rid of such
discontinuities, as well. This we pose to study in a future communication.\\
So far, we remain silent about the form of the potential. It is simply because, despite the most convenient
choice of the parameter, $f = 0.5$, it is still impossible to find an analytical solution for $\phi$, in view of
the solution (10). As a result, the form of the potential as a function of $\phi$ remains obscure. However, we
can plot the potential as a function of time, by choosing our second case, $n = 2$, for further simplification.
It is important to note that though the results sofar obtained, are independent of the value and signature of
$\lambda$, the form of the potential depends largely on it. In the following we make three such plots (taking
the help of "Manipulation" programme of Mathematica 6) to show how the form of the potential changes with
different values (starting from negative to large positive) of $\lambda$.
\begin{figure}
[ptb]
\begin{center}
\includegraphics[
height=2.034in, width=3.3797in] {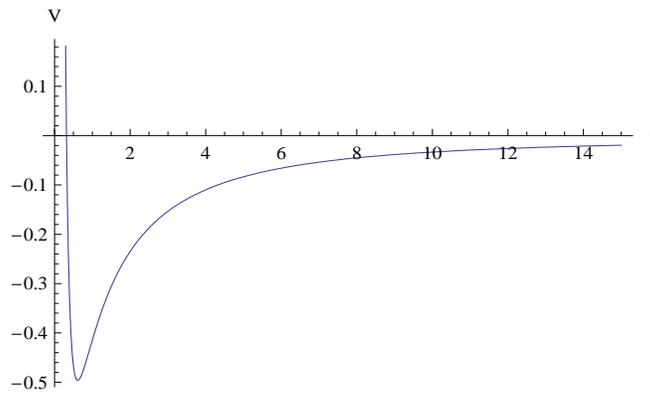} \caption{The form of the potential as a function of time for $\lambda
 \le -9$. Note that it remains negative throughout the evolution.}
\end{center}
\end{figure}

\begin{figure}
[ptb]
\begin{center}
\includegraphics[
height=2.034in, width=3.3797in] {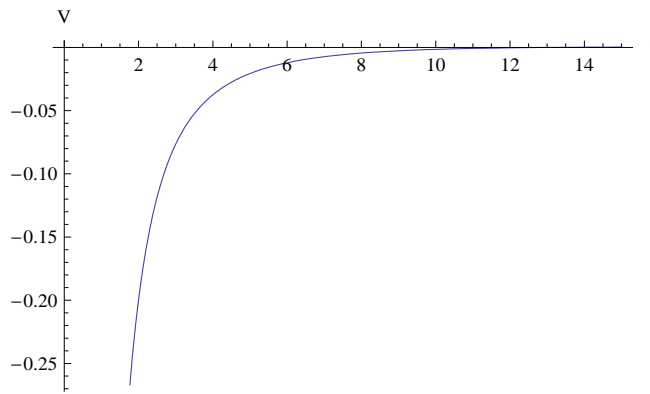} \caption{The form of the potential for $\lambda \approx 65$. Note that
it is zero at the present epoch but tends to grow in the future.}
\end{center}
\end{figure}

\begin{figure}
[ptb]
\begin{center}
\includegraphics[
height=2.034in, width=3.3797in] {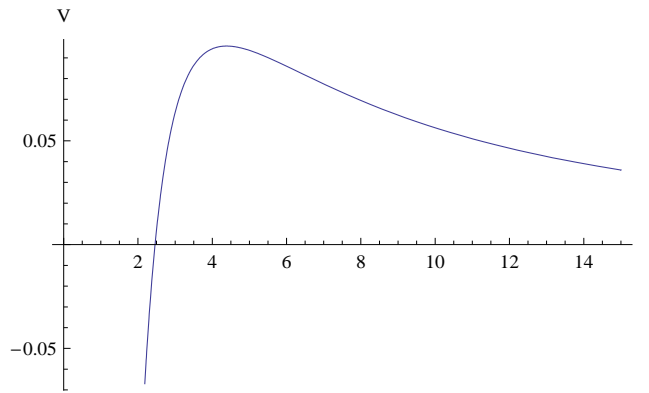} \caption{The form of the potential for $\lambda = 200$. Note that the
form is appreciably different.}
\end{center}
\end{figure}

\section{Concluding remarks}
Altogether we have obtained a late time transient crossing of the phantom divide line first from above and more
recently from below, starting from the inclusion of a Gauss-Bonnet-dilatonic scalar coupling term in the
standard Einstein-Hilbert action in four dimension. Since the crossing is transient, so such double crossing is
free from any sort of pathological behaviour both at the classical \cite{f} and at the quantum mechanical
\cite{g} levels. The striking feature of the model lyes in it's indistinguishability with the standard $\Lambda
CDM$ model, in terms of the luminosity-redshift and more precisely for distance modulus-redshift curves. To
identify between the two models we therefore require to observe dark energy equation of state $w_{de}$
independently. If $w_{de}$ is truly found to be dynamical and has really encountered a recent crossing of the
phantom divide line, then only we can definitely distinguish the standard $\Lambda CDM$ model with the present
one. It is highly interesting to learn that smooth transient crossing of the phantom divide line is allowed for
both negative and positive $(\lambda \lessgtr 0)$ type of Gauss-Bonnet-scalar interaction. Figures (7), (8) and
(9) also reveal that it is true even for different forms of the potential. Such transient crossing independent
of the value of $\lambda$ also signals that it might be possible to carry out the same treatment even for a
single scalar field model.

\textbf{Acknowledgement}:{Acknowledgement is due to Dipartimento di Scienze Fisiche, Universit$\grave{a}$  degli
studi di Napoli, Federico II, for their hospitality, TRIL (ICTP) for financial assistance and Prof. Claudio
Rubano for some illuminating discussion.}

\end{document}